\documentclass[superscriptaddress,nofootinbib,preprint]{revtex4}
\usepackage{graphicx}
\usepackage{amssymb,amsmath,amsfonts}
\usepackage[hypertex]{hyperref}

\def\lozenge{\boxit{\hbox to 1.5pt{\vrule height 1pt width 0pt \hfill}}}

\def\eg{{\it e.g.}}

\def\to{\rightarrow}

\newcommand{\vev}[1]{\langle {#1} \rangle}

\newcommand{\mP}{{\bar M}_P}

\newcommand{\dalam}{\raise-1mm\hbox{\large$\Box$}}

\newcommand{\beq}{\begin{equation}}
\newcommand{\eeq}{\end{equation}}

\begin{document}

\pagestyle{plain}

\hfill$\vcenter{ \hbox{\bf BNL-HET-07/16}
\hbox{\bf SLAC-PUB-12855}}$

\vskip 1cm
\title{On Direct Verification of Warped Hierarchy-and-Flavor Models}

\author{Hooman Davoudiasl}
\email{hooman@bnl.gov}
\affiliation{Department of Physics,
Brookhaven National Laboratory, Upton, NY 11973-5000, USA}
\author{{Thomas G. Rizzo}\footnote{Work supported in part
by the Department of Energy, Contract DE-AC02-76SF00515.}}
\email{rizzo@slac.stanford.edu}
\affiliation{Stanford Linear Accelerator Center, 2575 Sand Hill Rd.,
Menlo Park, CA  94025, USA}
\author{Amarjit Soni}
\email{soni@bnl.gov}
\affiliation{Department of Physics,
Brookhaven National Laboratory, Upton, NY 11973-5000, USA}


\begin{abstract}

We consider direct experimental verification of   
warped models, based on the Randall-Sundrum (RS) scenario, 
that explain gauge and flavor 
hierarchies, assuming that the gauge fields 
and fermions of the Standard Model (SM) propagate in the 5D bulk.  Most 
studies have focused on the bosonic Kaluza Klein (KK) signatures 
and indicate that discovering gauge KK modes is likely 
possible, yet challenging, while graviton KK modes are 
unlikely to be accessible at the LHC, even with a luminosity upgrade.  
We show that direct evidence for bulk 
SM fermions, {\it i.e.} their KK modes, 
is likely also beyond the reach of a luminosity-upgraded LHC.  Thus, 
neither the spin-2 KK graviton, the most distinct RS signal, 
nor the KK SM 
fermions, direct evidence for bulk flavor, 
seem to be within the reach of the LHC.  We then consider 
hadron colliders with $\sqrt{s} =$ 21, 28, and 60~TeV.  
We find that discovering the first KK modes of SM fermions and the graviton 
typically requires the Next Hadron Collider (NHC) with 
$\sqrt{s} \approx 60$~TeV and ${\cal O}(1)$~ab$^{-1}$ of 
integrated luminosity.  If the LHC yields hints of these warped models, 
establishing that Nature is 
described by them, or their 4D CFT duals,  
requires an NHC-class machine in the post-LHC experimental program.

\end{abstract}
\maketitle


\section{Introduction}

The Randall-Sundrum (RS) model \cite{Randall:1999ee} was originally proposed to resolve the 
hierarchy between the scales of weak and gravitational interactions, $m_W \sim 10^2$~GeV and 
$\mP \sim 10^{18}$~GeV, respectively.  The RS model is based on a truncated AdS$_5$ spacetime, bounded 
by two 4D Minkowski walls, often called UV (Planck) and IR (TeV) branes.  The curvature 
in 5D induces a warp factor in the metric which redshifts scales of order $\mP$ at the UV brane 
to scales of order $m_W$ at the IR brane.  Since the metric depends exponentially on the 5$^{th}$ 
coordinate, explaining $m_W/\mP \sim 10^{-16}$ does not require hierarchic parameters.  
The requisite brane-separation was shown to be easily accommodated early on \cite{Goldberger:1999uk}.  

Initially, it was assumed that all Standard Model 
(SM) fields reside at the IR brane.  
The most striking and distinct signature of this model would 
then be weak scale spin-2 Kaluza-Klein (KK) excitations 
of the graviton, appearing as resonances in high energy 
collisions \cite{Davoudiasl:1999jd}.  It was soon 
realized that resolving the hierarchy only required the Higgs to be 
localized near the IR brane \cite{Goldberger:1999wh} 
and SM gauge \cite{Davoudiasl:1999tf,Pomarol:1999ad} 
and fermion \cite{Grossman:1999ra}
fields could propagate in the 5D bulk.  
It was shown that placing the 
fermions in the bulk provides 
a natural mechanism for generation of SM fermion masses  
and also suppression of unwanted 4-fermion operators \cite{Gherghetta:2000qt}.  This 
is achieved by a mild modulation of bulk fermion masses that control 
the exponential localization of 
fermion zero modes.  As the Higgs is kept near the IR brane, small 4D Yukawa couplings 
are naturally obtained, if light flavor zero modes are UV brane localized.  
Given the correspondence between location in the bulk and scale in warped backgrounds, 
operators containing light flavors are suppressed by scales much larger than $m_W$. 

The above setup, an RS-type geometry with a flavored bulk, 
offers an attractive simultaneous resolution of hierarchy and flavor puzzles.  However, 
the experimental signals of these warped models are now much more elusive.  This is because 
spreading the gauge fields over the bulk and localizing the light fermions near the UV brane 
suppresses their couplings to IR-brane-localized KK modes, the main signatures of warped 
models.  Therefore, their production via and decay into light SM fermions and gauge fields 
are suppressed.  This feature is generic to warped models of hierarchy and flavor, 
largely independently of their details.

Recently, various studies have been performed to assess the prospects for discovering the 
new warped scenarios, given that the old set of signatures are now mostly 
inaccessible.  Precision data require the new KK states to be 
heavier than roughly 2-3~TeV \cite{Carena:2007ua}, even assuming new custodial 
symmetries \cite{custodial}.  Generally speaking, it has been shown that the most 
likely new state in these models to be discovered at the LHC 
is the first KK gluon \cite{KKgluon1,KKgluon2}.  The analysis 
of Refs.~\cite{KKgluon1,KKgluon2} suggests that KK gluons as 
heavy as 4~TeV will 
be within the reach of the LHC.  However, 
for the KK modes of the weak sector, 
the corresponding reach is in the 2-3~TeV range \cite{Agashe:2007ki}.  
For gauge KK masses in the above ranges, the graviton KK modes are most likely not 
accessible, even with an upgraded LHC luminosity \cite{KKgraviton1,KKgraviton2}.  Thus, typically, 
the gauge KK modes may be discovered, while the KK gravitons which are 
the most distinct RS-type signature will be out of reach, at the LHC.

In this work, we examine the discovery prospects for the KK modes of the SM 
fermion sector.  Observation of these states will provide direct evidence 
for the presence of SM fermions in the 5D bulk, a necessary ingredient 
of the warped flavor scenarios.  Briefly put, we find that with gauge KK masses set 
at 3~TeV, a currently acceptable value, and for generic zero-mode profiles that yield 
a realistic flavor hierarchy, the SM KK fermions are 
not accessible at the LHC, even after a luminosity upgrade.  Hence, we are faced with a 
situation in which the KK modes of the graviton and SM fermions are not discovered 
during the LHC program.  However, it seems reasonable to require direct observation 
of these KK states in order to establish an RS-type warped model as a theory of hierarchy 
and flavor.  

Given our conclusion that the LHC is unlikely 
to establish realistic warped models, we 
set out to determine the minimum 
requirements that a future 
machine needs to meet in order to make this task possible.  We take 3~TeV to be a reference 
mass for the lowest gauge KK mode.  This sets the mass scales of all other KK states, in the 
simplest generic warped models.  What we find is that, typically, in order to have 
firm evidence for the lowest KK states with spins 1/2, 1, and 2, we need center of mass 
energies ${\sqrt s}\approx 60$~TeV and integrated luminosities $L \sim 1$~ab$^{-1}$!  
This suggests that future {\it luminosity and energy} upgrades of the LHC will most likely 
be insufficient to verify all the essential features of realistic warped models.  Therefore, 
if these models do describe Nature, the LHC will likely find evidence for them.  However, 
the Next Hadron Collider (NHC), defined to have ${\sqrt s}\approx 60$~TeV 
and $L \sim 1$~ab$^{-1}$, must be part of the post-LHC 
experimental program aimed at establishing the underlying theory.  
We also show that if these states are not observed at the LHC the eventual 
reach of NHC for the gluon KK states is typically in excess of 10~TeV.  

In the next section we review key aspects of typical Warped Hierarchy-and-Flavor Models 
(WHFM).  In section 3, we study the prospects for verification of realistic WHFM at colliders 
and estimate the required parameters of the NHC.  Our conclusions
are presented in section 4.

\section{Warped Hierarchy-and-Flavor Models}

Here, we will briefly describe the generic properties of WHFM.  Much 
of what will follow is well-known from previous works and is mainly 
included to provide some background for our further discussions.

\subsection{General Features}

The RS metric is given by \cite{Randall:1999ee}
\beq
ds^2 = e^{-2 \sigma} \eta_{\mu\nu} dx^\mu dx^\nu - r_c^2 d\phi^2,
\label{metric}
\eeq
where $\sigma = k r_c |\phi|$, $k$ is the 5D curvature scale, 
$r_c$ is the radius of compactification, $-\pi\leq\phi\leq\pi$,
and a $\mathbb Z_2$ orbifolding of the 5$^{\rm th}$ dimension is assumed.  

To solve the hierarchy problem, the Higgs is assumed to be localized near the TeV-brane,
where the reduced metric ``warps" $\vev{H}_5 \sim \mP$ down to the weak scale:
$\vev{H}_4 = e^{-kr_c \pi} \vev{H}_5$.  For $kr_c \approx 11.3$, we then get
$\vev{H}_{\tt SM} \equiv \vev{H}_4 \sim 100$~GeV.  Originally, it was assumed that all SM
content resides at the IR-brane \cite{Randall:1999ee}.  
However, as the cutoff scale in the 4D
effective theory is also red-shifted to near the weak-scale, this would lead to
unsuppressed higher dimensional operators that result in large violations of experimental
bounds on various effects, such as those on flavor-changing neutral currents.  
This problem can be solved by realizing that points along the
warped 5$^{\rm th}$ dimension correspond to different effective 4D scales.
In particular, by localizing first and second generation fermions away from the IR-brane,
the effective scale that suppresses higher dimensional operators made up of these fields
is pushed to much higher scales \cite{Gherghetta:2000qt}.  
In the process of suppressing the dangerous operators,
this setup also leads to a natural mechanism for obtaining small fermion masses.

The above localization is achieved by introducing a 5D mass term in the bulk for
each fermion field $\Psi$ \cite{Grossman:1999ra}.  
Let $c \equiv m_\Psi/k$, where $m_\Psi$ is the 5D
mass of the fermion.  Each 5D fermion $\Psi$ has left- and right-handed
components $\Psi_{L,R}$ which can be expanded in KK modes
\beq
\Psi_{L,R}(x, \phi) = \sum_{n=0}^{\infty}\psi_{L,R}^{(n)}(x)
\frac{e^{2 \sigma}}{\sqrt{r_c}} f^{(n)}_{L,R}(\phi).
\label{KKmodes}
\eeq
The KK wavefunctions $f^{(n)}_{L,R}$ are orthonormalized
\beq
\int d\phi \, e^\sigma f^{(m)}_{L,R}\, f^{(n)}_{L,R}
= \delta_{m n}.
\label{orthonormal}
\eeq
One can then show that the $n\neq 0$ modes are given by \cite{Grossman:1999ra}

\beq
f^{(n)}_{L,R} = \frac{e^{\sigma/2}}{N^{L,R}_n}\, Z_{\frac{1}{2}\pm c} (z_n), 
\label{fKK}
\eeq
where the normalization $N^{L,R}_n$ is fixed by Eq.(\ref{orthonormal}); 
throughout our work, 
$Z_a = J_a + b_n Y_a$ denotes a linear 
combination of Bessel functions of order $a$.  In Eq.(\ref{fKK}),  
$z_n \equiv (m_n/k) e^\sigma$ and $m_n$ is the KK mass.  The zero-mode
wavefunction is given by
\beq
f^{(0)}_{L,R} = \frac{e^{\mp c \sigma}}{N^{L,R}_0},
\label{f0}
\eeq
with the normalization
\beq
N^{L,R}_0 = \left[\frac{e^{k r_c \pi(1\mp 2c)} - 1}{k r_c(1/2\mp c)}\right]^{1/2}.
\label{N0}
\eeq
Note that in our convention, the singlet (right-handed) and 
doublet (left-handed) zero mode wavefunctions are defined with 
opposite signs for mass parameters $c^S$ and $c^D$, respectively.  Hence, 
for example, UV-localization for the singlet and doublet zero modes 
correspond to $c^S < -1/2$ and $c^D > 1/2$, respectively.  

In the SM, all $SU(2)$ doublets are left-handed, while the singlets are
right-handed.  One can impose a ${\mathbb Z_2}$ parity on bulk
fermion fields so that only the doublets have left-handed zero modes and only
the singlets have right-handed zero modes. 
However, both the doublets and singlets have
left- and right-handed higher KK modes.  To project a particular 4D zero mode chirality, 
Neumann-like boundary conditions are chosen for the corresponding field at $\phi=0, \pi$; 
the other chirality will then obey Dirichlet boundary conditions.  For example, if 
$\Psi^D$ represents a weak SM doublet, we require 
\beq
(\partial_\phi + r_c m_\Psi)f^{(n)}_{L} = 0 \;\; ; \;\; f^{(n)}_{R} = 0 
\quad {\rm at}\;\; \phi = 0,\pi.
\label{bc}
\eeq 
The above choice results in a left-handed zero mode, given by Eq.(\ref{f0}).  We also find 
\beq
b_n^{L,R} = - \frac{J_{\pm(c - 1/2)}(z_n)}{Y_{\pm(c - 1/2)}(z_n)} 
\quad {\rm at}\;\; \phi = 0,\pi.
\label{bn}
\eeq
The above equations fix the wavefunctions and the mass eigenvalues, once $c$ is specified 
for any $SU(2)$ doublet $\Psi^D$; $(L,R)$ equations result in the same KK 
mass spectrum.  Similar 
equations can also be derived for singlets $\Psi^S$.
  
The above fermion profiles lead to a natural scheme for SM fermion masses 
\cite{Grossman:1999ra,Gherghetta:2000qt}.  We will assume that the Higgs 
is on the IR-brane; this is a very good approximation since the Higgs 
must be highly IR-localized.
Then, a typical Yukawa term in the 5D action will take the form
\beq
S^5_{\tt Y} = \int \!d^4x \; d\phi \sqrt{-g} \,\frac{\lambda_5}{k} H(x)
\Psi^D_L \Psi^S_R \delta(\phi-\pi),
\label{S5Y}
\eeq
where $\lambda_5 \sim 1$ is a dimensionless 5D Yukawa coupling and $\Psi^{D,S}$
are doublet left- and singlet right-handed 5D fermions, respectively.
After the rescaling $H \to e^{k r_c \pi} H$, the 4D action
resulting from Eq.(\ref{S5Y}) is
\beq
S^4_{\tt Y} = \int \!d^4x \; \sqrt{-g} \,\lambda_4 H(x)
\psi^{(D,0)}_L \psi^{(S,0)}_R + \ldots,
\label{S4Y}
\eeq
where the 4D Yukawa coupling for the corresponding
zero-mode SM fermion is given by \cite{Gherghetta:2000qt}
\beq
\lambda_4 = \frac{\lambda_5}{k r_c}\left[\frac{e^{(1-c^D+c^S)kr_c \pi}}
{N^{D,L}_0 N^{S,R}_0}\right],
\label{lam4}
\eeq
Thus, in the quark sector, there are, in general, 9 different 
values for $c^{D,S}$: 3 for the
doublets and 6 for the singlets.  One can see that the exponential form of the
effective Yukawa coupling $\lambda_4$ can accommodate a large
hierarchy of values without the need for introducing unnaturally small
5D parameters.  

With the fermions in the bulk, the gauge fields must also follow.  A 
5D gauge field $A_M$ has scalar and vector projections in 4D.  The 
scalar zero mode corresponding to $A_\phi$ is projected out using $\mathbb Z_2$ 
parity or a Dirichlet boundary condition.  As is well-known, 
the remaining projections $A_\mu$ 
can be expanded in KK modes as
\beq
A_\mu = \sum_{n=0}^\infty A_\mu^{(n)}(x) \frac{\chi^{(n)}(\phi)}{\sqrt{r_c}}.
\label{AKK}
\eeq
The gauge 
field KK wavefunctions are given by 
\cite{Davoudiasl:1999tf,Pomarol:1999ad}
\beq
\chi^{(n)}_A = \frac{e^{\sigma}}{N_n^A} Z_{1}(z_n)
\label{chiA}
\eeq
subject to the orthonormality condition
\beq
\int_{-\pi}^{\pi} d\phi \, \chi^{(m)}_A \chi^{(n)}_A 
= \delta_{m n}.
\label{ortho-gauge}
\eeq
The above equation fixes the normalization $N_n^A$ and 
Neumann boundary conditions at $\phi=0,\pi$ fix the wavefunctions and 
yield the KK masses.

The 5D action 
for the coupling of the bulk fermion 
$\Psi$ to the gauge field 
$A_M$ is given by
\beq
S_{\Psi A} = g_5\int \!d^4x \; d\phi V \left[V_l^M {\bar \Psi \gamma^l 
A_M \Psi}\right],
\label{PsiA}
\eeq
where $g_5$ is the 5D gauge coupling, 
$V$ is the determinant of the f\"{u}nfbein $V_l^M$, with 
$l=0,\ldots,3$, 
$V_\lambda^M = e^\sigma \delta^M_\lambda$, and $V_4^4 = -1$; 
$\gamma^l = (\gamma^\lambda, i \gamma^5)$.  
Dimensional reduction of the action (\ref{PsiA}) 
then yields the couplings of the fermion and gauge KK towers.  
With our conventions, the 4D gauge coupling is given by 
$g_4 = g_5/\sqrt{2 \pi r_c}$, which fixes the 
couplings of all the other modes.

For completeness, we also write down the wavefunction of the graviton KK 
modes \cite{Davoudiasl:1999jd}
\beq
\chi^{(n)}_G = \frac{e^{2\sigma}}{N_n^G} Z_{2}(z_n)
\label{chiG}
\eeq
which obey Neumann boundary conditions and are orthonormalized 
according to
\beq
\int_{0}^{\pi} d\phi \, e^{-2\sigma} \chi^{(m)}_G \chi^{(n)}_G 
= \delta_{m n}.
\label{ortho-grav}
\eeq
The graviton wavefunctions are fixed and the corresponding 
KK masses are obtained, via the boundary conditions, as before.  

WHFM are subject to various experimental constraints,
including those from precision
electroweak \cite{Carena:2007ua} and flavor data 
\cite{Agashe:2004cp,Agashe:2006iy}.  A number of models with a 
custodial $SU(2)_L\times SU(2)_R$ bulk symmetry have been
proposed to address these constraints 
\cite{Agashe:2003zs,Agashe:2004rs,Agashe:2006at}.  
In the following, we will
limit the scope of our study to bulk SM without specifying a
particular framework for such constraints.  We will not 
discuss the phenomenology of the extra exotica in these models, 
as they do not change the qualitative 
picture for the SM KK partners that
we present here.  This will suffice to
demonstrate our key observations.  
For a study of possible light exotic 
quarks in some warped scenarios 
see Ref.~\cite{Dennis:2007tv}. 

\subsection{Reference Parameters}

To comply with precision constraints, we will choose the 
mass of the first KK mode of gauge fields to be 3~TeV \cite{Carena:2007ua}.  
This implies that we have assumed extra custodial symmetries, as explained in the 
above.  We ignore brane localized kinetic terms for various bulk 
fields, as they are most naturally loop suppressed.  Then, the RS 
geometry fixes the ratios of all KK masses.  In what follows, 
we will also ignore potential mixing among various KK modes that 
can induce small shifts in their masses.  These considerations do not 
change our main conclusions regarding discovery reach for WHFM, at colliders.  
The masses of various KK modes are given by 
\beq
m_n = x_n \, k e^{-kr_c \pi},
\label{mKK}
\eeq
where for gauge fields $x_n = 2.45, 5.56, 8.70, \ldots$ and 
for the graviton $x^G_n = 3.83, 7.02, 10.17,\ldots$.  

The values of KK masses for SM fermions depends on the bulk mass parameter $c$.  
In the fermion sector, we will only discuss the reach for SM KK quarks, since we will 
concentrate on multi-TeV hadron colliders.   
Here, we choose $c^D \approx - c^S \approx 0.6$ for {\it light} quarks.  
This choice results in masses of order 10-100~MeV for 
${\cal O}(1)$ bulk Yukawa couplings.  This range roughly covers the  
quark flavors $(u,d,s)$ which constitute the dominant quark initial 
states for collider production of new physics.   
It turns out that the KK modes of all quarks, except for 
the singlet top quark, are roughly 
degenerate in mass with KK modes of the gauge fields.  To get a reasonable 
top quark mass, we choose $c^D_t \approx c^S_t \approx 0.4$, giving a 
singlet top first KK mass roughly 1.5 times the first gauge KK mass.

To assess the relative significance of various production channels, we 
must know the relevant couplings that enter the calculations (an earlier 
qualitative discussion in a somewhat different context can be found in 
Ref.\cite{Davoudiasl:2007zx}).  Again, 
we only discuss typical values to keep our discussion more general and 
less parameter-specific.  We adopt the notation $g^l_{mn}$ to 
denote the coupling of the $l^{th}$ gauge KK mode to fermions of the 
$m^{th}$ and $n^{th}$ KK levels; $g^0_{00} \equiv g_{SM}$.  We will focus on 
the gluon and quark sectors, so $g_{SM} = g_s$, the strong coupling constant.

First, we will consider single production of KK quarks.  This cannot proceed via 
fusion of quark and gluon zero modes, since this vertex is zero by orthogonality of 
fermion KK wavefunctions.  Next, is the possibility of production in association with a 
zero mode quark.  The gluon-mediated diagram again gives zero, by orthogonality.  
However, KK-gluon-mediation gives a non-zero result.  Given that 
KK gluons are IR-localized, the only feasible channels 
involve the third generation zero modes.  
We find that the coupling $g^1_{00}\sim g_s/5$, for light quarks.  
Note that the initial states 
cannot be gluons, by orthogonality of the gluon KK modes.  
For $(t, b)_L$ and 
$t_R$, we get $g^1_{01} \sim \sqrt{2\pi}g_s$.  Hence, production in these channels 
is roughly proportional to $(\sqrt{2 \pi}g_s^2/5)^2\sim g_s^4/4$.  Given that the 
KK modes of the singlet top are about 1.5 heavier than the doublets, 
the most promising 
channel is single production of a third generation 
doublet KK mode in association with a $t$ or a $b$.

Next, let us examine pair production of KK quarks.  Gluon mediated production 
can come from zero mode quark and gluon initial states; each of these amplitudes is 
proportional to $g_s^2$.  There is also a KK gluon mediated 
channel.  Here, the amplitude is roughly proportional to $g_s^2$; this is
due to an approximate cancellation of volume suppression 
and enhancement at the two vertices 
for this process.  Therefore, we see that pair-production has a number of SM-strength 
channels available to it, unlike the single production.  This can offset the 
kinematic suppression from producing two heavy states.  In fact we will later see that 
pair-production will dominate single-production with our typical choices of parameters.

Finally, since we will also discuss discovery reach for KK 
gravitons, we will briefly review their relevant couplings.  As is well-known, 
light quark zero modes have a negligible coupling to KK gravitons.  This is because 
of the extreme IR-localization of these KK gravitons, compared with KK gauge fields.  Thus, 
in generic WHFM, only the coupling $C^{ A A G }_{ 0 0 n }$ 
of two gluons to the KK graviton  is important for its collider 
production \cite{KKgraviton1,KKgraviton2}.  
In units of the graviton zero mode coupling, $1/\mP$, we have 
\cite{Davoudiasl:2000wi} 
\beq
C^{A A G}_{0 0 n} = 
e^{k r_c \pi}
\frac{2 \left[1 - J_0 \left(x_n^G \right) \right]}{k r_c \pi 
\left(x_n^G \right)^2 |J_2 \left(x_n^G \right)|},
\label{ggG}
\eeq
where the first few $x^G_n$ were given above.

\section{Experimental Prospects for Verification of WHFM}

Given the discussion in the previous section, the first thing we need to do is to demonstrate 
that single (associated) production of KK fermions together with their corresponding zero modes is 
relatively suppressed in comparison to the production of KK fermion pairs. We will concentrate on 
the quark and gluon sectors.  This reaction is dominated by the subprocess 
$q\bar q \to g^{(1)}\to q^{(1)}\bar q^{(0)}+h.c.$, 
where the first gluon KK mode $g^{(1)}$ is somewhat off-shell. (We assume the width to mass 
ratio of this gluon KK to be 1/6 in our analysis, following  
Ref.~\cite{KKgluon1}.) To be concrete, we focus our attention on 
the excitations of the left-handed third generation quark 
doublet $Q_3=(t,b)_L^T$ since it has a large coupling to $g^{(1)}$, as discussed above. 

\begin{figure}[htbp]
\centerline{
\includegraphics[width=7.5cm,angle=90]{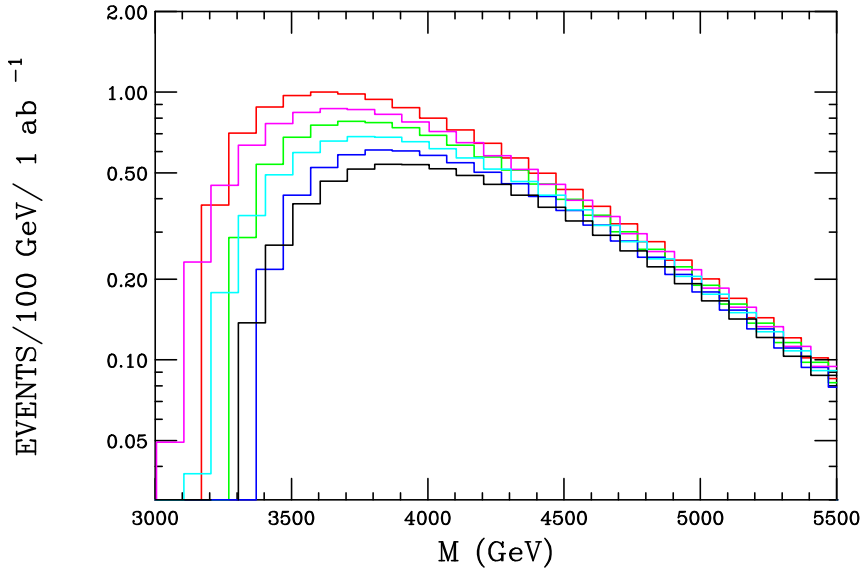}}
\vspace*{0.1cm}
\caption{Rates for the associated production of first $t_L$ and $b_L$ KK excitations together with 
their corresponding zero modes at the LHC; the results for $t(b)_L$ correspond to  
to the higher (lower) member of each histogram pair, respectively. Here the first gluon KK mass 
is fixed at 3~TeV and we show the results as a function of the fermion pair mass. 
The three sets of histograms assume 
that the $b_L$ KK mass is 100 $(200, 300)$~GeV heavier than the 
gauge KK whereas the threshold for $t_L$ is somewhat higher  
due to the larger zero mode top mass.}
\label{fig1}
\end{figure}

\begin{figure}[htbp]
\centerline{
\includegraphics[width=7.5cm,angle=90]{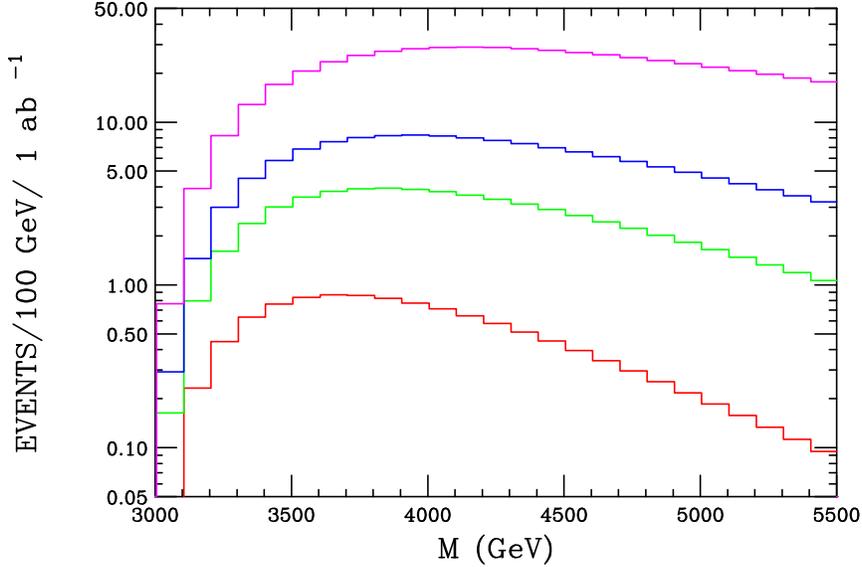}}
\vspace*{0.1cm}
\caption{Same as the last figure but now for different values of $\sqrt s $ and taking the first gluon 
KK and fermion KK masses to be degenerate at 3~TeV. From bottom to top the histograms correspond to 
$\sqrt s=14$, 21, 28 and 60 TeV, respectively.}
\label{fig2}
\end{figure}

For the LHC, the results of these calculations for single production 
can be seen in Fig.~\ref{fig1} while 
generalization to higher energy colliders can be seen 
in Fig.~\ref{fig2}. Note that no cuts or branching 
fractions have been included in these calculations, 
in order to avoid a model-specific analysis. 
These rates will go down if we assume a more 
massive gluon KK state. In this work, we 
will generally assume that ${\cal O}(100)$ events 
will be sufficient to establish the discovery of a KK mode.  
The results in Figs. (\ref{fig1}) and (\ref{fig2}) 
represent the case with the product of the 
entrance and exit channel couplings equal to $g_s^2$, the SM value.  
However, from our discussion in the previous section, 
we expect that a more realistic value, for these processes, is $g_s^2/2$, suppressing the 
production by factor of about 4.   
The rates, as can be seen from the figures, 
are not very impressive, even with this optimistic assumption. 
It is clear that this process will be 
unobservable at the LHC even with a luminosity upgrade.  
Further, note that the cross section 
only increases by a factor of order $\sim 20$ 
in the peak region when going from $\sqrt s=$14 to 60 TeV.

\begin{figure}[htbp]
\centerline{
\includegraphics[width=7.5cm,angle=90]{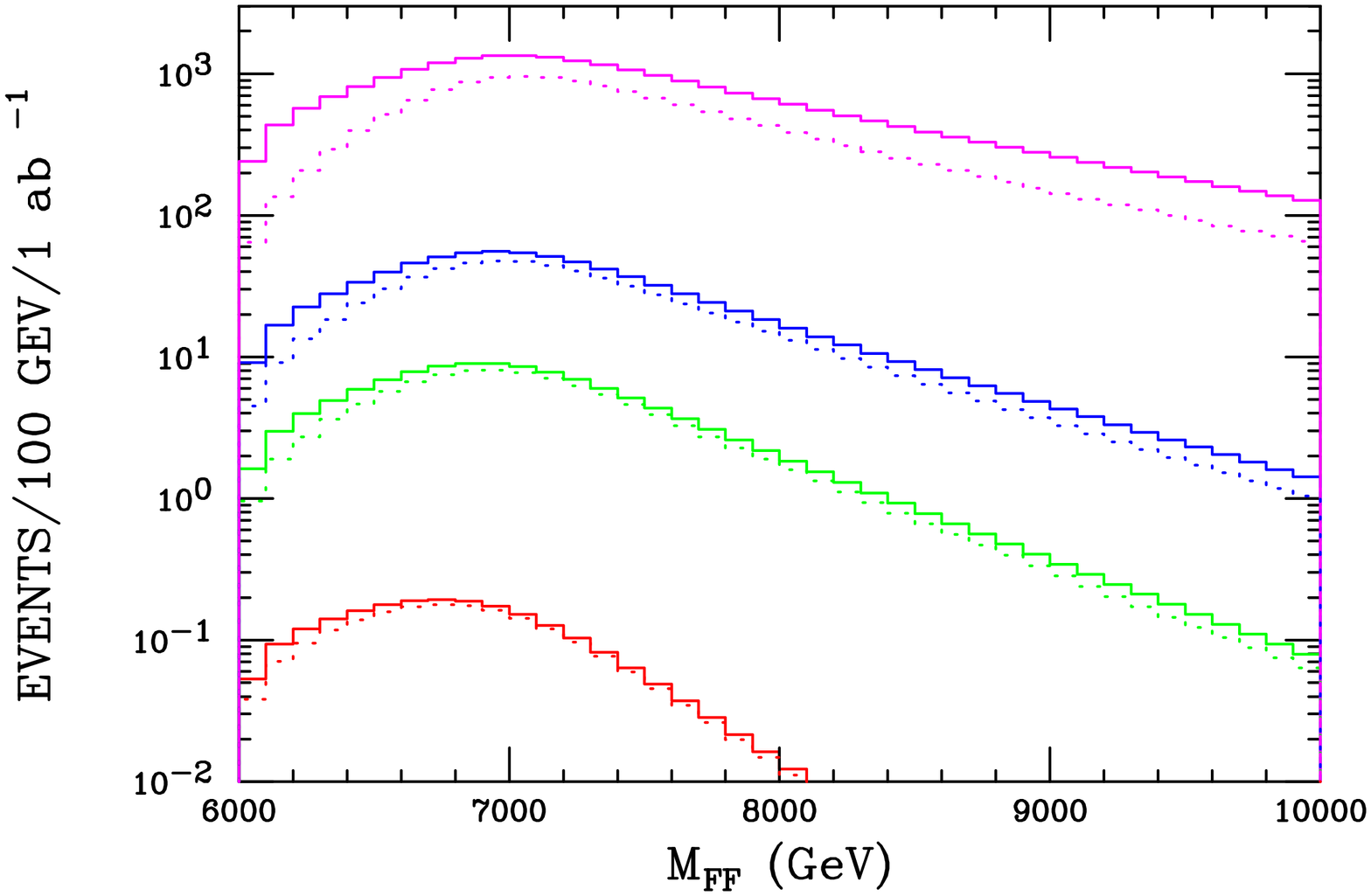}}
\vspace*{0.1cm}
\caption{Fermion KK pair production rate at various collider energies as a function of the fermion pair 
invariant mass assuming that the first gluon and fermion
KK states are degenerate with a mass of 3 TeV. From bottom to top the histograms correspond to $\sqrt s=14$, 21, 
28 and 60 TeV, respectively. No cuts have been applied. The dotted histogram shows the result of only including 
zero mode gluon exchange.}
\label{fig3}
\end{figure}

\begin{figure}[htbp]
\centerline{
\includegraphics[width=7.5cm,angle=90]{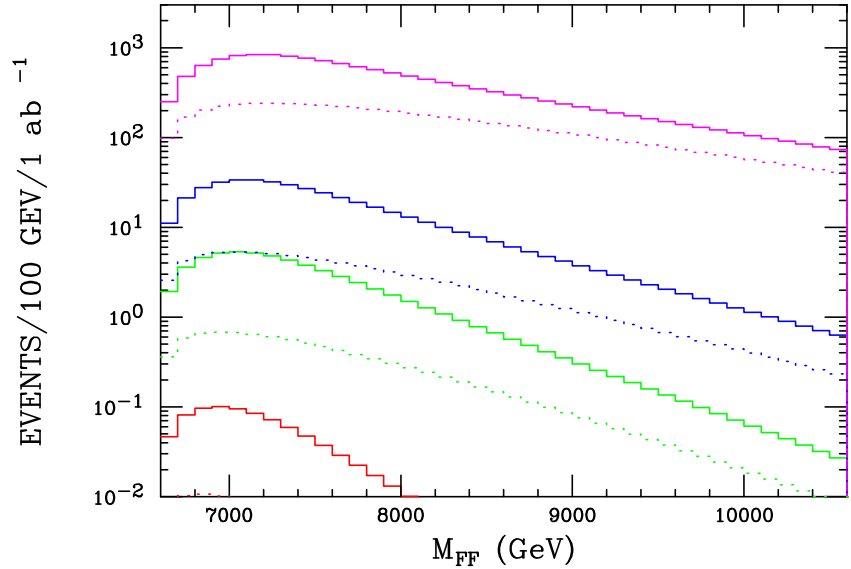}}
\vspace*{0.1cm}
\caption{Same as the previous figure but now assuming that 
the KK fermion is $10\%$ more massive than 
the first gauge KK state.}
\label{fig4}
\end{figure}

We now turn to the possibility of fermion KK pair production which can arise from both $q\bar q$ and $gg$ 
initial states as discussed above and is mediated by the entire gluon KK tower, including the zero mode. (Note  
that in our calculations we include only the first three gluon KK excitations as well as the SM gluon.) The 
results of these calculations are shown in Fig.~\ref{fig3} and \ref{fig4} with no cuts or branching fractions 
applied. Again we see that for the LHC the rates are far too small to be useful but they grow quite rapidly as 
the collider energy increases; without much effort we see that $\sqrt s=28$ GeV is perhaps the bare minimum 
requirement to observe these KK states and an even higher value 
is likely to be necessary if efficiencies and branching 
fractions are suitably accounted for.

We will now discuss the likely dominant decay channels for 
KK quarks, without entering into a detailed analysis.  
We will consider the gauge eigen-basis picture, for simplicity.  
Given that the KK quarks are somewhat heavier than 
the gluon KK modes, $q_{KK} \to q \, g_{KK}$, with $q$ a zero-mode quark, is a possible 
decay channel; let us denote this possibility as channel A.  
Another potentially important decay mode is through the brane-localized Yukawa 
coupling to the Higgs: $q^i_{KK} \to H \, q^j_{KK}$ where the final quark KK mode is lighter 
than the one in the initial sate; we will 
refer to this as channel B.  Channel C in the following will refer to $q^i_{KK} \to H \, q^j$ 
with $q^j$ denoting another zero-mode quark. 
(The weak-sector analog of channel A will be suppressed by the weak coupling constant so 
we will ignore it in the discussion which follows.) 
Note that here if $q^i$ is a weak 
doublet then $q^j$ is a weak singlet, and vice versa.  The subsequent decay of the gluon KK state 
in channel A produces the final state $q\, t{\bar t}$.  In channel B, depending on 
whether $q^j_{KK}$ decays through channels A or C, we get either $H\,q\,t{\bar t}$ or 
$H\, H\, q$, respectively as a final state.  

The couplings in channels A and C are controlled by the overlap of the 
zero mode quark with the IR brane states, $g_{KK}$ or $H$.  For channel B 
this coupling is ${\cal O}(1)$, when allowed, since it involves three IR-brane states 
(CFT composites, in the dual picture).   
For example, if a singlet quark is much more UV-localized than 
its doublet counterpart, the doublet KK mode will decay through channel A, 
whereas the singlet KK mode will decay mostly through channel B (since it 
is more massive than the doublet KK mode).  With our reference values, 
the singlet and doublet KK modes of light flavors 
are taken to be degenerate, which essentially eliminates channel B.  However, 
this is not typically expected to be the case in a realistic setup, 
where doublet and singlet quarks have differing values of profile 
parameter $c$.  We have also ignored the effect of mixing in the 
mass eigen-basis, where one expects that the 
degenerate KK modes are split by off-diagonal 
Yukawa couplings \cite{Agashe:2006iy}.  Exactly 
what decay modes will dominate for each quark KK state depends on the 
$c$ values and possible mixing angles.  Nonetheless, given the above 
three channels, the typical final states corresponding to 
the decay of the pair-produced quark KK modes are expected to be 
given by  $2\times [q\, t{\bar t}]$, $2\times [H\,q\,t{\bar t}]$, or 
$2\times [H\,H\,q]$.  Needless to say, these are complicated final states and 
require further study regarding their reconstruction and possible backgrounds.

A complete verification of WHFM, based on the RS picture, 
would require the observation of the first graviton KK state.  
This is known to be difficult at LHC energies if the gauge KK 
mass is 3 TeV; the graviton KK mass in this case is 
$\simeq 4.7$ TeV. A promising search mode is to look for 
the process $gg\to G^{(1)} \to Z_LZ_L$ \cite{KKgraviton2} which 
is rather clean; here $Z_L$ denotes a longitudinal $Z$ which 
is IR-localized. 
The SM background arises from the conventional 
tree-level process $q\bar q \to ZZ$ via 
$t-$ and $u-$ channel diagrams which is highly peaked in the 
forward and backward directions and is reducible 
by strong rapidity cuts. Note that as the $\sqrt s$ of the 
collider increases the average collision energy also increases. 
This leads to a stronger peaking of the SM $ZZ$ background in 
both the forward and backward directions. Since the decay 
products of the $Z$'s essentially follow those of their original 
parent particle a tightening of the rapidity
cuts will be necessary as $\sqrt s$ increases to 
maintain a reasonable signal to background ratio.  
In performing these calculations, given our assumptions, 
the only free parameter is the ratio $k/\mP$.

\begin{figure}[htbp]
\centerline{
\includegraphics[width=7.5cm,angle=90]{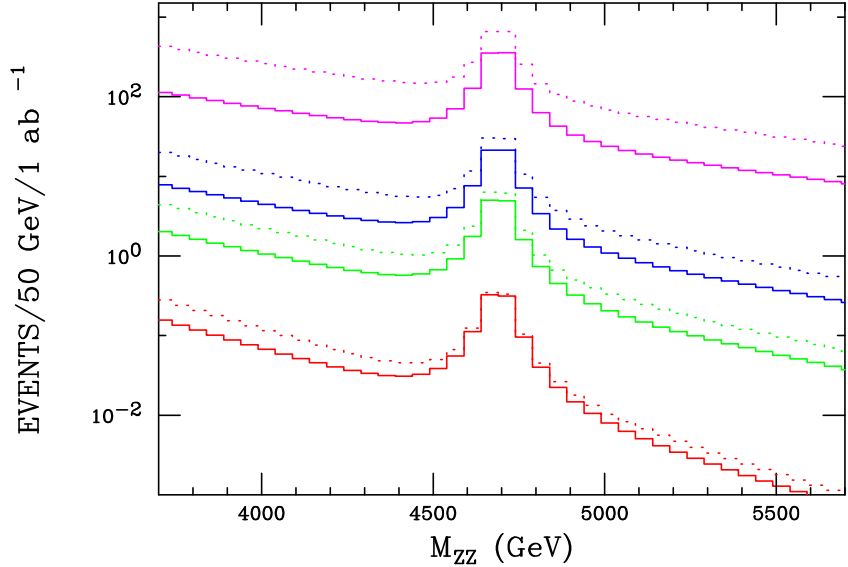}}
\vspace*{0.1cm}
\caption{Production rate for the first graviton KK excitation decaying into two $Z$ bosons assuming a rapidity 
cut $|y|<2(1)$ on the $Z$'s corresponding to the dotted(solid) histograms. The histograms correspond, from bottom 
to top, to collider energies of $\sqrt s=14$, 21, 28 and 60 TeV, respectively, $Z$ branching fractions are not 
included and $k/\mP=0.5$ has been assumed.}
\label{fig5}
\end{figure}

\begin{figure}[htbp]
\centerline{
\includegraphics[width=7.5cm,angle=90]{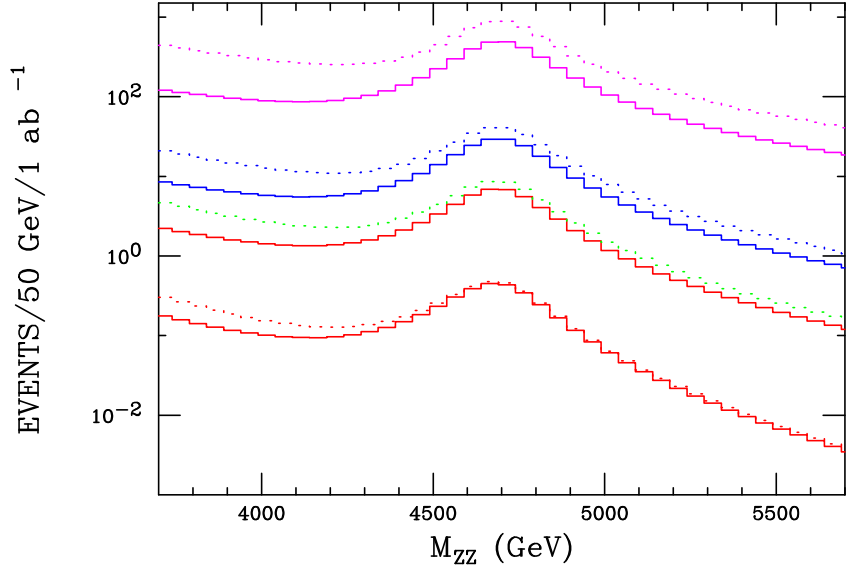}}
\vspace*{0.1cm}
\caption{Same as the previous figure but now with $k/\mP=1.0$.}
\label{fig6}
\end{figure}

\begin{figure}[htbp]
\centerline{
\includegraphics[width=7.5cm,angle=90]{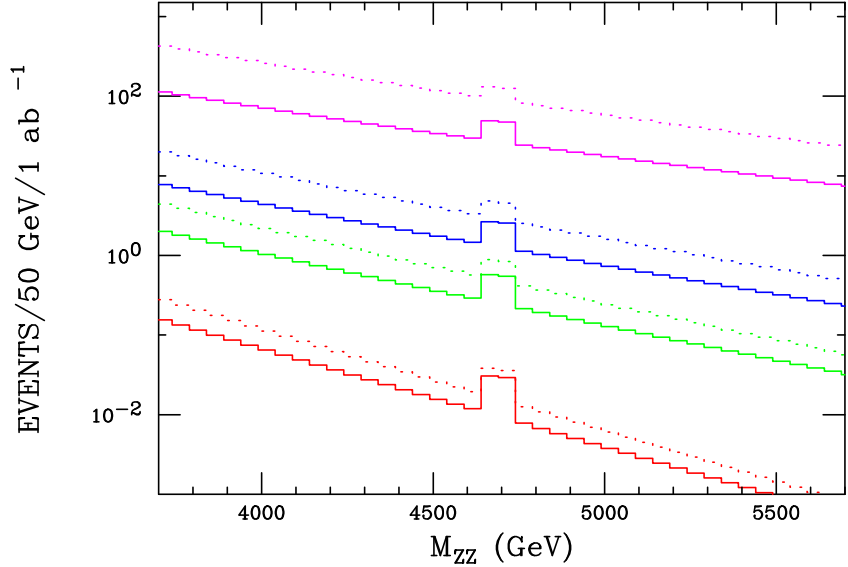}}
\vspace*{0.1cm}
\caption{Same as the previous figure but now with $k/\mP=0.1$.}
\label{fig7}
\end{figure}

Figs.~\ref{fig5},~\ref{fig6} and ~\ref{fig7} show the results of these calculations for three different values 
of the ratio $k/\mP=0.5,1.0$ and 0.1, respectively; as expected, in all cases the event rate is far too low 
at the LHC to be observable. For $k/\mP=0.5$, the graviton KK is a reasonably well-defined resonance structure 
which grows quite wide (narrow) when this same ratio equals 1 (0.1). For a fixed rapidity cut, 
the signal over the 
background is is seen to diminish as the center of mass energy of the 
collider grows larger. If we identify the 
graviton through the decay $ZZ \to jj \ell^+ \ell^-$, which 
has a branching fraction of a few percent, it is clear 
that a $\sqrt s \sim 60$ TeV collider will be necessary 
to observe this state. Here, we ignore detection issues related 
to having collimated jets from highly boosted $Z$'s.  A realistic 
jet resolution could make these conclusions less optimistic.  However, 
we refrain from making assumptions about the detector 
capabilities and data analysis methods of future NHC experiments.  
Note that if $k/\mP$ is sufficiently 
small even this large collider energy will be insufficient 
as to discover the graviton KK resonance as it becomes quite narrow. 

It is also of some interest to examine the possibility of 
searching for gluon KK states via pair production now that we 
are considering higher energy colliders. The production of 
resonant single gluon KK states is made difficult since it can 
only occur through the light initial state partons which have 
suppressed couplings to the KK gluons.  
The possibility of pair 
production avoids these issues as it arises from both $gg$ 
as well as $q\bar q$ initial states and occurs through the exchange 
of the complete gluon KK tower; we include the first 
three gluon KK states in the calculations below in addition to the SM gluon 
zero mode.  Once produced, gluon KK states 
almost uniquely decay to top quark pairs, in typical 
scenarios \cite{KKgluon1, KKgluon2}.  
Since 2 KK gluon states are made, the final 
state consists of two pairs of top quarks. 

\begin{figure}[htbp]
\centerline{
\includegraphics[width=7.5cm,angle=90]{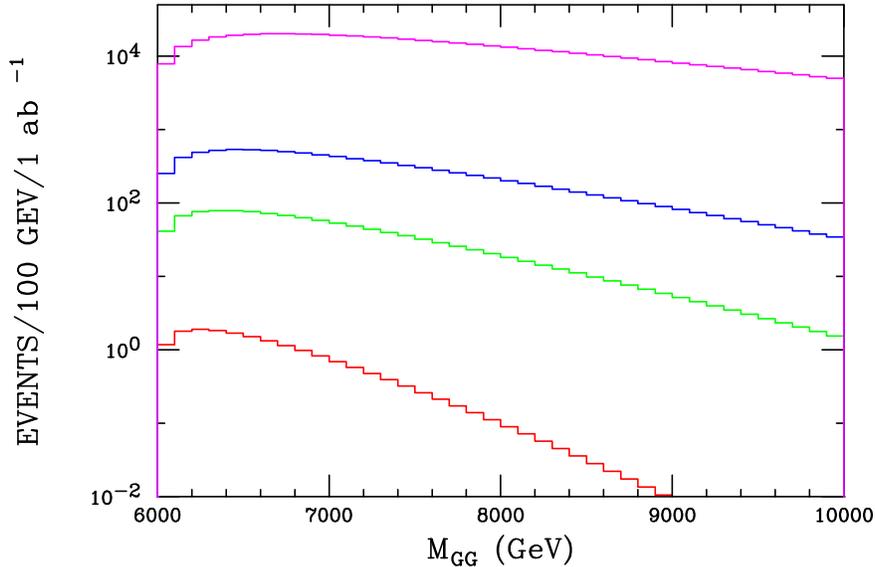}}
\vspace*{0.1cm}
\caption{Same as Fig.~\ref{fig3} except now for the pair production of the lightest gluon KK state.}
\label{fig8}
\end{figure}

Fig.~\ref{fig8} shows the results of these calculations for 
the same collider energies as above, {\it i.e.}, $\sqrt s=14$, 21, 28 
and 60 TeV. As can be seen here, these signal rates can be quite 
significant once we go to energies substantially above that of the LHC. In 
order to determine how significant the resulting signal from these rates 
is we need to have an estimate of the SM background.  
We will not consider this background here and 
only provide the production rate. 

\begin{figure}[htbp]
\centerline{
\includegraphics[width=7.5cm,angle=90]{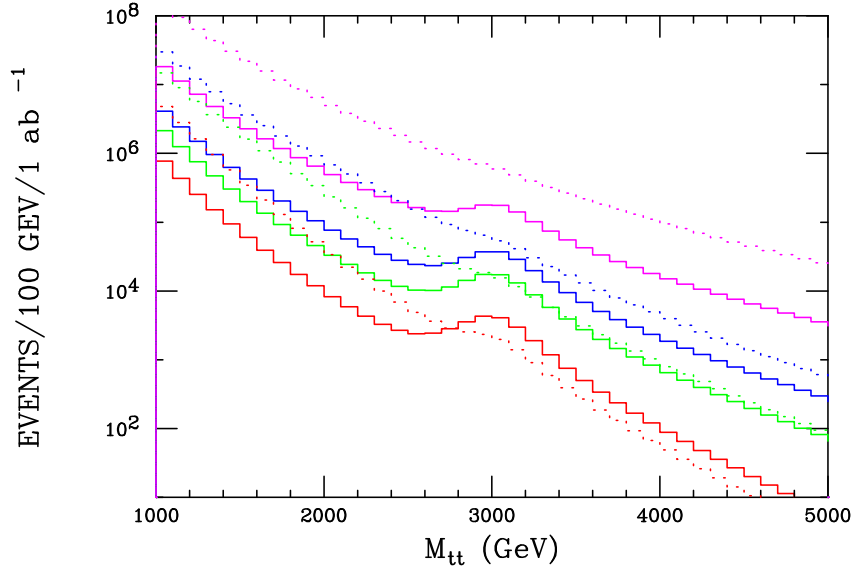}}
\vspace*{0.1cm}
\caption{The resonant production rate for the first 
gluon KK state in the $t\bar t$ channel assuming $|y|<2.5$ for
the dotted case and $|y|<1$ for the solid one.
From bottom to top the histograms 
correspond to $\sqrt s=14$, 21, 28 and 60 TeV respectively. 
No efficiencies or branching fractions are included.}
\label{fig9}
\end{figure}

The $q\bar q \to g^{(1)} \to t\bar t$ gluon KK channel 
essentially sets the discovery reach for the RS scenario at 
hadron colliders \cite{KKgluon1, KKgluon2}.  At 
the LHC, a 3 TeV KK gluon should be visible 
above the usual top quark pair SM background, with suitable rapidity cuts, after branching 
fractions and tagging efficiencies 
are accounted for. Unfortunately, as in the graviton KK $ZZ$ 
mode above, as the center of mass energy 
of the collider increases the forward/backward peaked SM 
$gg,q\bar q \to t\bar t$ process grows quite rapidly relative to the KK gluon 
signal for fixed rapidity cuts.  
(Recall that at these energies the decay 
products of the top quark will essentially follow the original top 
flight direction.)  This result can be seen quite 
explicitly in Fig.~\ref{fig9}. Here we see that the obvious gluon KK peak 
structure slowly disappears with increasing collider energy.

\begin{figure}[htbp]
\centerline{
\includegraphics[width=7.5cm,angle=90]{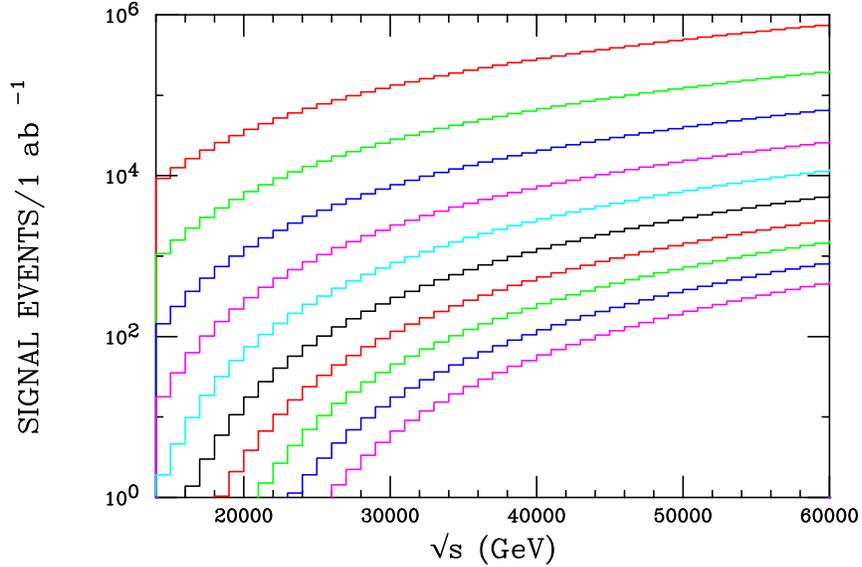}}
\vspace*{0.1cm}
\caption{Signal rate for a possible gluon KK resonance as a function of the collider energy employing 
the cuts described in the text. Branching 
fractions and efficiencies have been neglected. From top to bottom the results are 
shown for gluon KK masses in the range from 3 to 12 TeV 
in steps of 1 TeV.}
\label{fig10}
\end{figure}

If we want to ensure a significant signal to background ratio for gluon KK states 
at higher energy colliders we need to tighten our rapidity 
cuts from the usual `central detector' requirements, $|y|<2.5$. In our analysis 
below we will assume that $|y|<1$ to increase the S/B ratio. 
Furthermore we will define the signal region to be in the $t\bar t$ invariant mass 
range within $\pm \Gamma_{KK}$, the gluon KK width, of the 
gluon KK mass with $\Gamma_{KK}/M_{KK}=1/6$ assumed in our analysis below. Fig.~\ref{fig10} 
shows the resulting signal rates following this 
procedure for a range of KK gluon masses as a function of the collider energy; branching 
fractions and efficiencies have been ignored in 
obtaining these results.

\begin{figure}[htbp]
\centerline{
\includegraphics[width=7.5cm,angle=90]{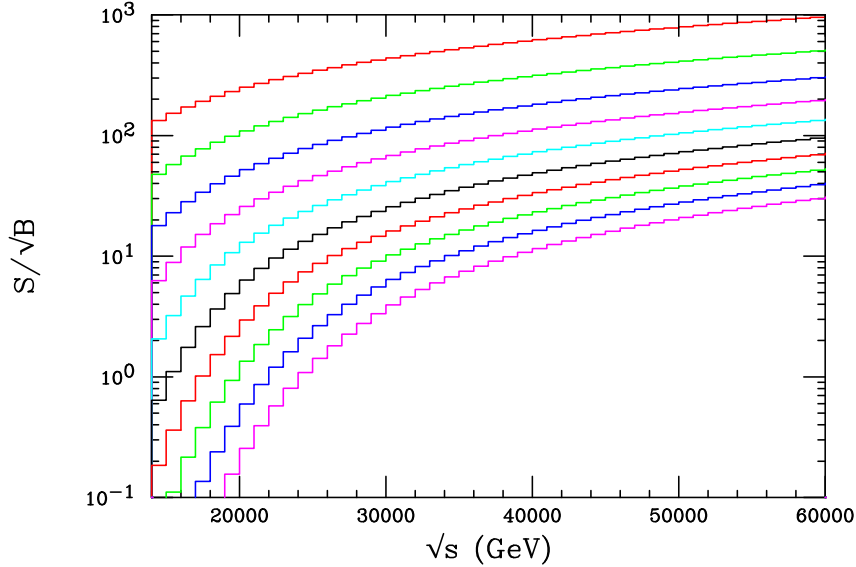}}
\vspace*{0.1cm}
\caption{Same as the previous figure but now showing the signal significance. Again, branching 
fractions and efficiencies have been neglected.}
\label{fig11}
\end{figure}

Fig.~\ref{fig11} shows the signal significance for gluon KK production for a range of 
masses as a function of the hadron collider center 
of mass energy. In a more realistic calculation which includes top quark branching fractions $f$ 
and $b$-tagging efficiencies $\epsilon$, the results 
shown here must be scaled by $\sqrt {f \epsilon} \sim 0.15$ \cite{KKgluon1}. 
Taking this factor into account we see that, \eg, at a $\sqrt s= 21\, (28, 60)$ TeV collider 
KK gluon masses as large as 5.5 (7, 12) TeV may become accessible. 
One could say that this covers the entire `natural' parameter space for WHFM.

\section{Conclusions}

Amongst models of 
physics beyond the Standard Model, WHFM are rather unique to
the extent that they are capable of providing
a simultaneous natural resolution of 
two important puzzles, namely the hierarchy and the flavor problems.
In particular, this is in sharp contrast
to supersymmetry, which is otherwise 
an extremely interesting theoretical construct.

Thus, it is clearly important to establish the 
requirements for direct experimental 
verification of WHFM.  These more recent developments  
of the original RS model \cite{Randall:1999ee}, which explain 
gauge and flavor hierarchies in one framework, have KK 
particles in the few TeV range.
Low energy precision tests suggest that gauge KK masses in 
WHFM are likely to be heavier than about 3 TeV.
Perhaps the most compelling and unique signature of these
models is the spin-2 KK graviton. Unfortunately, it seems at the 
LHC, even with an upgraded luminosity, KK gravitons,
expected here to lie above 4.7 TeV, 
are very challenging to observe and likely 
inaccessible. Prospects for KK gluons up to masses around
4 TeV seem brighter at the LHC. Thus it may well be
that some early indication of the underlying RS idea
may find support at the LHC.  However, though considerable work exists in the
literature on warped {\it bosonic} KK modes,
not much has been done for the SM fermion counterparts.  The  
observation of these {\it fermionic} KK modes can in essence 
be taken as direct experimental evidence for warped bulk flavor.   

With that perspective in mind, in this work, we set out
to provide an exploratory study of the parameters needed
for the next generation of machines that could provide 
significant experimental support
for the  WHFM, and especially for generation of 
flavor through bulk localization.
We concentrate on hadron colliders only, 
as they are expected to yield the largest kinematic reach.  For definiteness,
throughout our numerical study here, we take the
lightest gauge KK mass to be 3 TeV; SM fermionic KK modes are always as 
massive or heavier.

First, we studied single KK fermion production.
The most promising candidate in this regard is the KK mode
of the third generation doublet produced in association
with a $t$ or a $b$ quark.   We found that at the LHC the 
prospects for finding this KK fermion through single production 
are rather grim. In fact, we showed that even a 28 TeV machine is only
likely to see at most a handful of candidate events of this category.
The prospects improve significantly for a 60 TeV machine
wherein a few tens of events are possible.

Pair production of KK fermions typically seems to have over an order
of magnitude larger cross section compared to the single-production, for 
large $\sqrt{s}$.
For pair production, a 28 TeV machine can give several
tens of events and 60 TeV produces hundreds 
of such candidate events.  One may expect that a 
sample of this latter size is needed for a reliable verification 
of the bulk flavor scenarios, after cuts and efficiencies are taken 
into account;  we do not delve into these issues in this 
exploratory work.

We have revisited the earlier study of the KK graviton  
through the ``gold plated" $ZZ$ mode \cite{KKgraviton2}. 
We find that for this unique channel, 
only a 60 TeV machine can provide
${\cal O}(100)$ events over a 
plausible range of $k/\mP$ values.  Thus, it seems that verification 
of the WHFM, in the sense of directly measuring the properties of signature 
KK modes, requires what we refer to as the 
Next Hadron Collider (NHC) with $\sqrt{s} \approx 60$~TeV and 
${\cal O} (1)$~ab$^{-1}$ of integrated luminosity.  
We have also extended earlier studies of the resonance production
of the first KK gluon state and its detection through  
a $t \bar t$ pair. We find that colliders with energies
14, 21, 28 and 60 TeV can allow detection of the first KK gluon 
up to masses 4, 5.5, 7, and 12 TeV, respectively.     

We hence conclude that 
an NHC-class machine must be an integral part of the 
high energy experimental program if hints of WHFM are 
discovered at the LHC.  The same 
conclusion holds for 4D models that explain 
hierarchy and flavor in a similar fashion and constitute  
dual dynamical scenarios, according to 
AdS/CFT correspondence \cite{Maldacena:1997re,ArkaniHamed:2000ds}.

\acknowledgments

H.D. and A.S. are supported in part by the DOE grant
DE-AC02-98CH10886 (BNL).

\end{document}